\documentclass[11pt,modern]{aastex}

\def\GRS  {GRS~1915+105}
\def\Cyg  {Cyg~X-3}

\def\p    {\phantom{0}}
\def\q    {\phantom{00}}
\def\r    {\phantom{000}}

\def\kms  {km~s$^{-1}$}

\def\masy {mas~y$^{-1}$}

\def\uas  {\ifmmode {\mu{\rm as}}\else{$\mu$as}\fi}
\def\deg  {\ifmmode {^\circ}\else {$^\circ$}\fi}
\def\porm {\ifmmode {\pm}\else {$\pm$}\fi}
\def\chisqpdf {\ifmmode {\chi^2_{\rm pdf}}\else {$\chi^2_{\rm pdf}$}\fi}
\def\chisq    {\ifmmode {\chi^2}\else {$\chi^2$}\fi}
\def\Msun {M$_\odot$}

\def\HI   {H~{\small I}}

\def\eg   {e.g.,~}

\def\d    {\ifmmode {{\rlap{.}}^\circ}\else {${\rlap{.}}^\circ$}\fi}
\def\s    {\ifmmode {{\rlap{.}}^s}\else {${\rlap{.}}^s$}\fi}
\def\as   {\ifmmode {{\rlap{.}}^{''}}\else {${\rlap{.}}^{''}$}\fi}

\newbox\grsign \setbox\grsign=\hbox{$>$} \newdimen\grdimen \grdimen=\ht\grsign
\newbox\laxbox \newbox\gaxbox
\setbox\gaxbox=\hbox{\raise.5ex\hbox{$>$}\llap
     {\lower.5ex\hbox{$\sim$}}}\ht1=\grdimen\dp1=0pt
\setbox\laxbox=\hbox{\raise.5ex\hbox{$<$}\llap
     {\lower.5ex\hbox{$\sim$}}}\ht2=\grdimen\dp2=0pt

\def\lax{\mathrel{\copy\laxbox}}

\def\pa    {\ifmmode {\psi} \else {$\psi$}\fi}
\def\rPpm  {\ifmmode {r_{\Ro,\To}} \else {$r_{Ro,\To}$}\fi}

\def\vlsr  {\ifmmode {V_{\rm LSR}}\else {$V_{\rm LSR}$}\fi}
\def\vlsrr {\ifmmode {v^r_{\rm LSR}}\else {$v^r_{\rm LSR}$}\fi}
\def\vhelio{\ifmmode {v_{Helio}}\else {$v_{Helio}$}\fi}

\def\ura   {\ifmmode {\mu_\alpha}\else {$\mu_\alpha$}\fi}
\def\udec  {\ifmmode {\mu_\delta}\else {$\mu_\delta$}\fi}

\def\ul    {\ifmmode {\mu_l}\else {$\mu_l$}\fi}
\def\ub    {\ifmmode {\mu_b}\else {$\mu_b$}\fi}

\def\uml   {\ifmmode {v_{gr}}\else {$v_{gr}$}\fi}
\def\umb   {\ifmmode {v_b}\else {$v_b$}\fi}
\def\vsrad {\ifmmode {v_{rad}}\else {$v_{rad}$}\fi}

\def\upl   {\ifmmode {v^p_{gr}}\else {$v^p_{gr}$}\fi}
\def\upb   {\ifmmode {v^p_b}\else {$v^p_b$}\fi}
\def\vprad {\ifmmode {v^p_{rad}}\else {$v^p_{rad}$}\fi}

\def\Vo    {\ifmmode {V^{Std}_\odot}\else {$V^{Std}_\odot$}\fi}
\def\Uo    {\ifmmode {U^{Std}_\odot}\else {$U^{Std}_\odot$}\fi}
\def\Wo    {\ifmmode {W^{Std}_\odot}\else {$W^{Std}_\odot$}\fi}
\def\VH    {\ifmmode {V^H_\odot}\else {$V^H_\odot$}\fi}
\def\UH    {\ifmmode {U^H_\odot}\else {$U^H_\odot$}\fi}
\def\WH    {\ifmmode {W^H_\odot}\else {$W^H_\odot$}\fi}
\def\V     {\ifmmode {V_\odot}\else {$V_\odot$}\fi}
\def\U     {\ifmmode {U_\odot}\else {$U_\odot$}\fi}
\def\W     {\ifmmode {W_\odot}\else {$W_\odot$}\fi}

\def\Vs    {\ifmmode {V_s}\else {$V_s$}\fi}
\def\Us    {\ifmmode {U_s}\else {$U_s$}\fi}
\def\Ws    {\ifmmode {W_s}\else {$W_s$}\fi}

\def\Vsbar {\ifmmode {\overline{V_s}}\else {$\overline{V_s}$}\fi}
\def\Usbar {\ifmmode {\overline{U_s}}\else {$\overline{U_s}$}\fi}
\def\Wsbar {\ifmmode {\overline{W_s}}\else {$\overline{W_s}$}\fi}

\def\aone  {\ifmmode {a_1}\else {$a_1$}\fi}
\def\atwo  {\ifmmode {a_2}\else {$a_2$}\fi}
\def\athr  {\ifmmode {a_3}\else {$a_3$}\fi}

\def\pars  {\ifmmode{\pi_s}\else{$\pi_s$}\fi}

\def\Ts    {\ifmmode{\Theta_s}\else{$\Theta_s$}\fi}
\def\Tdot  {\ifmmode{d\Theta\over dR}\else{$d\Theta\over dR$}\fi}

\def\Rp    {\ifmmode{R_p}\else{$R_p$}\fi}

\def\To    {\ifmmode{\Theta_0}\else{$\Theta_0$}\fi}
\def\Ro    {\ifmmode{R_0}\else{$R_0$}\fi}

\def\Dp    {\ifmmode{d_p}\else{$d_p$}\fi}
\def\Zsun  {\ifmmode {Z_\odot}\else {$Z_\odot$}\fi}
\def\Ytilt {\ifmmode{\psi_Y}\else{$\psi_Y$}\fi}
\def\Xtilt {\ifmmode{\psi_X}\else{$\psi_X$}\fi}

\def\ZIAU  {\ifmmode {Z_{IAU}}\else {$Z_{IAU}$}\fi}

\usepackage{listings}
\usepackage{appendix}
\usepackage{lineno}
\usepackage{float}
\usepackage[caption = false]{subfig}
\usepackage{geometry}
\shorttitle{Cyg~X3 and GRS 1915 Distances }
\shortauthors{Reid \& Miller-Jones}

\begin{document}

\pagestyle{plain}

\title{\bf On the Distances to the X-ray Binaries Cygnus~X-3 and \GRS}

\author[0000-0001-7223-754X]{M. J. Reid}
\affiliation{Center for Astrophysics~$\vert$~Harvard \& Smithsonian,
   60 Garden Street, Cambridge, MA 02138, USA}
\author[0000-0003-3124-2814]{J. C. A. Miller-Jones}
\affiliation{International Centre for Radio Astronomy Research, 
   Curtin University, GPO Box U1987, Perth, WA 6845, Australia}

\begin{abstract}
In this paper we significantly improve estimates of distance to the X-ray binary
systems \Cyg\ and \GRS.  We report a highly accurate trigonometric parallax
measurement for \Cyg\ using the VLBA at 43 GHz, placing the source at a distance
of $9.67^{+0.53}_{-0.48}$\,kpc.  We also use Galactic proper motions and
line-of-sight radial velocity measurements to determine 3-dimensional (3D)
kinematic distances to both systems, under the assumption that they have low
peculiar velocities. This yields distances of $8.95\pm0.96$\,kpc for \Cyg\
and $9.4\pm0.6~({\rm statistical})\pm0.8~({\rm systematic})$ for \GRS.
The good agreement between parallax and 3D kinematic distances validates the
assumption of low peculiar velocities, and hence small natal kicks, for both of
the systems.  For a source with a low peculiar velocity, given its parallax distance,
\Cyg\ should have a \vlsr\ near $-64\pm5$ \kms.
Our measurements imply a slightly higher inclination angle, and
hence lower black hole mass for \GRS\ than found from previous work by
\citet{Reid:14grs} and strengthen arguments from X-ray polarization that
\Cyg\ would be an ultraluminous X-ray source if viewed face-on.

\end{abstract}

\section{Introduction}

Knowledge of the distance to an astronomical source is fundamental for
estimating its true nature, including its mass and luminosity.  The case of
the high-mass X-ray binary Cyg~X-1 is an excellent example.  It was the
first binary suggested to include a black hole, based on its periodic
velocity excursions and the lack of an observable companion
\citep{Webster:72,Bolton:72}.  However, for nearly 40 years, one could
not be certain whether the companion was a black hole or a neutron star,
since distance estimates ranged by more than a factor of two, from about
1.1 to 2.5 kpc \citep[see, \eg][]{Caballero-Nieves:09}, and at the lower
end of the range of distances companion masses could be below about 5 \Msun.
This problem was resolved by \citet{Reid:11}
and \citet{Miller-Jones:21}, using the Very Long Baseline Array (VLBA)
to observe the radio emission from the compact companion and measure a
trigonometric parallax relative to background quasars, with the latter study yielding a distance of
$2.22^{+0.18}_{-0.17}$ kpc.  This firmly established that the Cyg~X-1
system contains a black hole and a massive young star.

However, accurate parallaxes for more distant X-ray binaries have been
hard to obtain.  In particular, two well-studied X-ray binaries, \GRS\ and Cyg~X-3,
have large distance uncertainties, which limit our understanding of their
nature.  \citet{Reid:14grs} observed \GRS\ with the VLBA and measured a
{\it relative} parallax to a nearby (in both projection and distance) water
maser associated with a massive, young star.  Combining the relative parallax
of \GRS\ with the absolute parallax of the maser \citep{Wu:14}, and prior
constraints on distance based on models of jet kinematics, resulted in a
distance estimate for \GRS\ of $8.6^{+2.0}_{-1.6}$ kpc and led to
an estimate of its compact companion mass of $12.4^{+2.0}_{-1.8}$ \Msun.
Previous distance estimates had only indirectly constrained it to be larger
than about 6 kpc, inferred from \HI\ absorption \citep{Mirabel:94},
and smaller than 12.5 kpc, based on the ratio of apparent
speeds of the approaching and receding jets \citep{Fender:99}.

Determining the distance to Cyg~X-3 has been less successful than for \GRS.
\citet{Dickey:83} noted that there is absorption from interstellar \HI\ toward
Cyg~X-3 from Local Standard of Rest velocities, \vlsr, of zero to at least $-70$ \kms,
suggesting a lower limit for its distance of $>11.6\times(\Ro/10~{\rm kpc})$,
where $\Ro$ is the distance to the Galactic center.
\citet{Predehl:00} compared the angular extent of the X-ray halo of
Cyg~X-3 with the time delay of X-rays scattered by intervening dust and
estimated a distance of $9^{+4}_{-2}$ kpc.  \citet{Ling:09} re-analyzed
the X-ray data and, assuming that the scattering occurs in the
Cyg OB2 association of young stars at 1.7 kpc, estimated a distance of Cyg~X-3
of $7.2^{+0.3}_{-0.5}$ kpc.   However, allowing the Cyg OB2 distance to be between
1.38 and 1.82 kpc places Cyg~X-3 between 3.4 and 9.3 kpc distant.

In this paper we present a very accurate trigonometric parallax for the
high-mass X-ray binary Cyg~X-3, as well as an independent estimate of its
distance using 3-dimensional (3D) kinematics.
For the micro-quasar \GRS, which already has a trigonometric parallax measurement,
we also provide an independent estimate of distance using 3D kinematics.
Finally, we carefully examine the fundamental assumption of kinematic distances --
that the sources have only small to moderate ($\lax20$ \kms) non-circular motions --
as this has strong implications for how some compact stars can form, since 
it requires small natal ``kicks.''

\section{Estimating Distance with 3D Motions}\label{3D_method}

\citet{Reid:22} analyzed the use of proper motions in addition to line-of-sight
velocities to obtain 3D kinematic distance estimates, concluding that
they had great potential for sources more distant than about 8 kpc.
3D kinematic distance estimates compare the full observed velocity vector
to a model of Galactic rotation, with distance as an adjustable parameter.
The fundamental Galactic parameters -- the distance to the Galactic center (\Ro),
the circular rotation speed of the Sun (\To), and the rotation curve of the Milky
Way -- are now known to near 1\% accuracy \citep[see][for details]{Reid:19,2019Sci...365..664D,2020ApJ...892...39R,2021A&A...647A..59G}.
For distant sources, measurements of proper motion can require orders of magnitude less
precision compared to parallax measurements for similar fractional distance uncertainty.
Thus, 3D kinematic distances offer an opportunity to refine distance estimates for
sources that follow Galactic rotation.

For the Galactic model needed for kinematic distance estimates, we assume the
values of $\Ro=8.15$ kpc and $\To=236$ \kms\ from \citet{Reid:19} and their
Solar Motion parameters ($\U=10.6, \V=10.7, \W=7.6$) \kms.  For the rotation curve of the
Milky Way, we adopt that of \citet[][documented in their appendix B]{Reid:19}.
This rotation curve model follows the 2-parameter ``universal'' formulation of
\citet{Persic:96} and was obtained by fitting 147 maser sources with Galactocentric
radii between 4 and 15 kpc using measured 3D motions and ``gold standard'' parallax
distances.  Following  \citet{Reid:22}, we estimate 3D kinematic distances by 
forming likelihoods as a function of distance for three components of motion:
the velocity with respect to the Local Standard of Rest, \vlsr,
the proper motion in Galactic longitude, \ul, and
the proper motion in Galactic latitude, \ub.
Assuming a flat prior on distance, the product of these likelihoods gives
the combined posterior distribution function (PDF) for distance.

\section{\GRS}

The proper motion of \GRS\ has been measured relative to compact extra-galactic
sources by \citet{Dhawan:07} and \citet{Reid:14grs} using the VLBA of the National Radio Astronomy Observatory\footnote{The National Radio Astronomy Observatory is a facility of the National Science Foundation operated under cooperative agreement by Associated Universities, Inc.}. 
\citet{Dhawan:07} observed predominantly at 8.4 GHz between 1996 and 2006 and achieved
single-epoch astrometric precision of $\approx1$ mas.  They measured the eastward
and northward motions to be $\ura=-2.86\pm0.07$ \masy\ and $\udec=-6.20\pm0.09$ \masy.
Independently, \citet{Reid:14grs} observed at 22 GHz between 2008 and 2013 and with improved
astrometric techniques, including using ``geodetic blocks'' to measure and remove residual
tropospheric delays \citep{Reid:09}, and achieved single-epoch precision of $\approx0.2$ mas,
yielding $\ura=-3.19\pm0.03$ \masy\ and $\udec=-6.24\pm0.05$ \masy.
The variance-weighted average is $\ura=-3.14\pm0.03$ \masy, $\udec=-6.23\pm0.04$ \masy,
which converts to motions in Galactic longitude and latitude of
$\ul=-6.98\pm0.05$ \masy, $\ub=-0.12\pm0.01$ \masy.
\citet{Reid:14grs} recalibrated the data of \citet{Steeghs:13} and estimated the heliocentric
line-of-sight velocity $\gamma=+12.3\pm1.0$ \kms, corresponding to a $\vlsr=30.4$ \kms.

\begin{figure}
\epsscale{0.77} 
\plotone{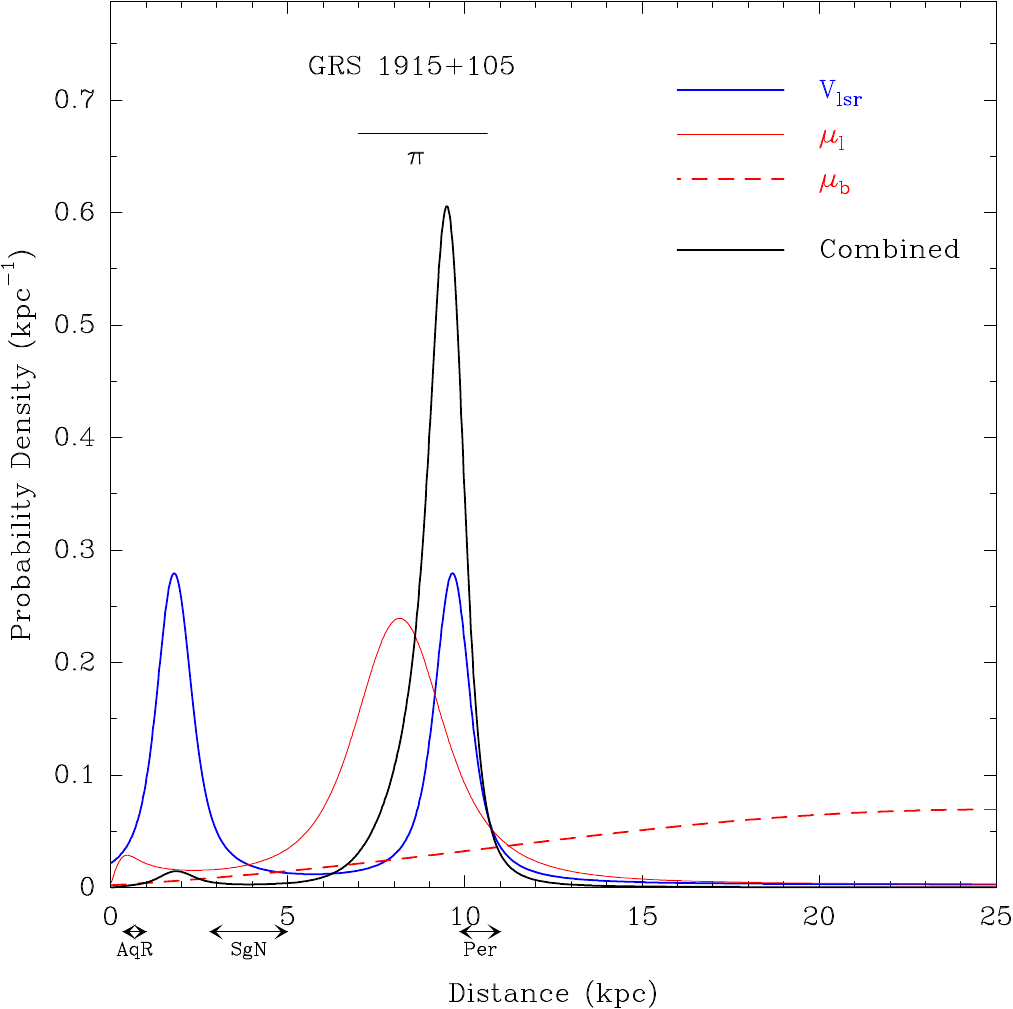}
\caption{\small Likelihoods for three components of motion, \vlsr\ in ({\it blue}),
  Galactic longitude ({\it red solid line}) and latitude ({\it red dashed line}),
  as a function of distance for \GRS.
  The product of the three likelihoods is shown in {\it black}, indicating
  a distance of $9.4 \pm 0.6~({\rm statistical}) \pm 0.8~({\rm systematic})$ kpc.
  The parallax-based distance from \citet{Reid:14grs} is indicated with the $\pi$
  symbol and its 68\% confidence range is given above it.  The range of distances
  for spiral arms along the line-of-sight are indicated below the distance axis.
    }
\label{fig:GRS1915}
\end{figure}

Figure \ref{fig:GRS1915} displays the likelihood functions for the three components
of motion of \GRS.  The line-of-sight velocity (\vlsr) gives maximum likelihood
distances of 1.80 or 9.65 kpc; the Galactic longitude proper motion favors
a distance of 8.15 kpc; and the latitude motion weakly constrains the distance
(favoring a large value).  The combined 3D kinematic distance estimate for \GRS\ is
$9.4 \pm 0.6~({\rm statistical}) \pm 0.8~({\rm systematic})$ kpc,
where the statistical uncertainty is a Gaussian $1\sigma$ width of
the combined PDF and the systematic uncertainty is half of the
separation of the peaks of the line-of-sight (\vlsr) and longitude motion (\ul)
likelihoods.  

Using the same fundamental parameters of the Milky Way adopted
for the model for the 3D kinematic distance, the non-circular (peculiar)
motion components for \GRS\ are $(U,V,W) = (18\pm2, 8\pm22, 2\pm2$) \kms,
where $U$ is toward the Galactic center, $V$ is in the direction of
Galactic rotation, and $W$ is toward the North Galactic Pole.
Thus, the magnitude of the peculiar motion of \GRS\ is fairly small
($\sim20$ \kms).

\section{Cyg~X-3}

\subsection{Trigonometric Parallax}

Previous attempts to measure the parallax of Cyg~X-3 used the VLBA under program BM343.
Those observations at 12 GHz employed background quasars for calibration which
were separated from Cyg~X-3 by $\approx3\deg$.
Owing to this large separation, these observations yielded only a marginal parallax detection.
The lack of compact quasars near Cyg~X-3 is a result of strong scattering from
interstellar electrons over a few degrees on the sky toward the Cygnus X region.
Such scattering increases the apparent angular size of radio sources, making them heavily
resolved on long interferometer baselines.  Since scattering angles decrease as the
inverse-square of observing frequency, in order to minimize scatter broadening and find a
closer background source, we surveyed continuum radio sources within $2\deg$ of Cyg~X-3
with the VLBA at 43 GHz and found one, J2033+4000, which was relatively compact and
separated by only $1\deg$ from Cyg~X-3.

In VLBA program BR212, we observed Cyg~X-3 and J2033+4000 at 43 GHz at eight
epochs spanning one year.  The calibrator, J2033+4000, was resolved on the longest
baselines, and we only used seven antennas (stations codes: BR, FD, KP, LA, NL, OV, PT)
with a maximum baseline length of 2300 km.  We ``nodded'' the array between the
two sources, changing sources every 20 sec, in order to transfer phase from
J2033+4000 to Cyg~X-3 within the interferometer coherence time, limited by rapid
fluctuations in water vapor.  We also calibrated the slowly varying (hours time-scale)
changes in total water vapor above each antenna by observing ``geodetic-like''
blocks of quasars at 24 GHz across the sky.  These and other calibration methods
are described in detail in \citet{Reid:09}.  The data were correlated using the
VLBA DiFX software correlator \citep{Deller:11}, and analyzed using the 
Astronomical Image Processing System \citep{Greisen:2003}.

\begin{figure}[h]
\epsscale{0.65} 
\plotone{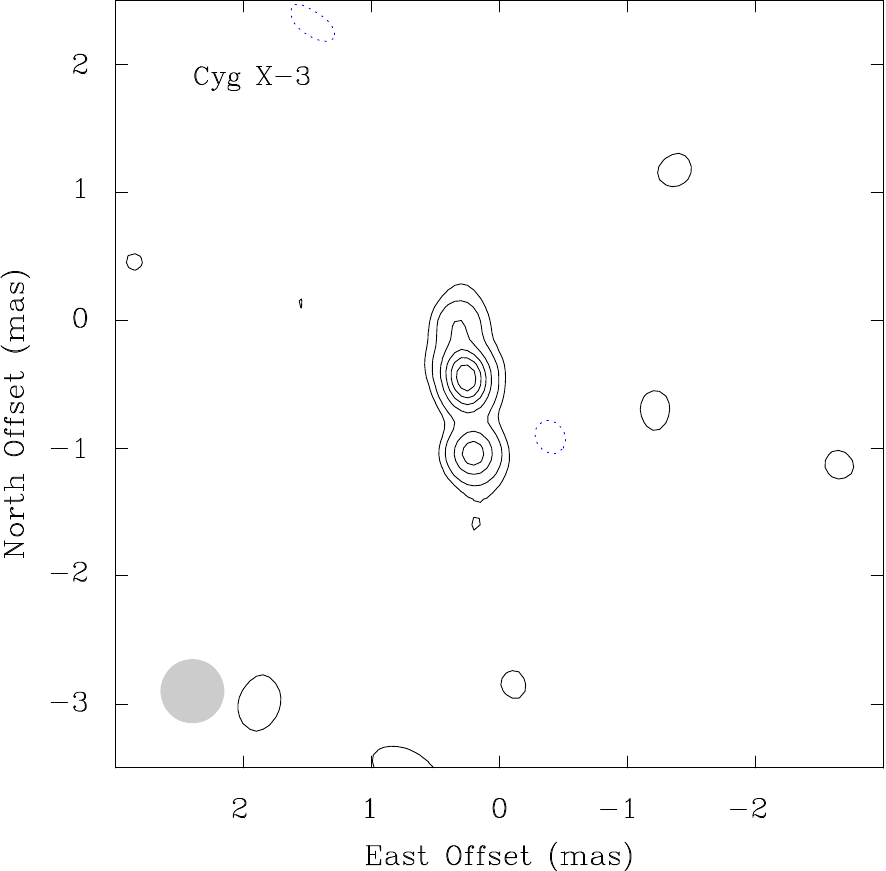}
\caption{\small VLBA contour map of \Cyg\ at 43 GHz on 2017 May 14 when the source was weakest.  Contour levels are -0.5, 0.5, 1.0, 2.0, 3.0, 4.0, 5.0 mJy. A 0.5 mas FWHM beam is shown as the shaded disk at the bottom left. The brightest component, near $(0.2,-0.5)$ mas, was used for fitting the parallax. 
    }
\label{fig:image}
\end{figure}

Fig.~\ref{fig:image} shows a representative image of Cyg~X-3 from observations
on 2017 May 14.   The northern bright spot was clearly visible at all epochs
and served as the astrometric point for the parallax measurement.   
Table \ref{table:cygx3_data} gives the dates of the
observation, and measured positions and brightnesses for Cyg~X-3 obtained
by fitting a Gaussian brightness distribution to the northern spot.

\begin{deluxetable}{crrr}
\tablecolumns{4} \tablewidth{0pc} 
\tablecaption{Parallax Data for Cyg~X-3}
\tablehead {
  \colhead{Date} &\colhead{East Offset} &\colhead{North Offset}  &\colhead{Brightness}\\
  \colhead{}     &\colhead{(mas)}       &\colhead{(mas)}         &\colhead{(mJy\,beam$^{-1}$)}      
           }
\startdata
2016.806   &$1.495\pm0.003$     &$1.702\pm0.005$  &$18.5\pm0.2$  \cr
2016.836   &$1.425\pm0.004$     &$1.548\pm0.007$  &$19.0\pm0.4$  \cr
2017.308   &$0.397\pm0.001$     &$-0.072\pm0.002$ &$40.2\pm0.3$  \cr
2017.325   &$0.346\pm0.006$     &$-0.159\pm0.007$ &$15.3\pm0.4$  \cr
2017.344   &$0.291\pm0.004$     &$-0.319\pm0.005$ &$13.6\pm0.2$  \cr
2017.368   &$0.266\pm0.007$     &$-0.441\pm0.010$ &$ 6.7\pm0.2$  \cr
2017.815   &$-1.126\pm0.003$    &$-2.146\pm0.004$ &$16.8\pm0.2$  \cr
2017.847   &$-1.185\pm0.005$    &$-2.180\pm0.007$ &$14.2\pm0.4$  \cr
\enddata
\tablecomments{\footnotesize Column 1 gives the date of the observation.
  Columns 2 and 3 give the measured position offsets of Cyg~X-3
  relative to J2033+4000, after removing a constant angular difference
  assuming J2000 coordinates of (20:32:25.76955,+40:57:27.8820) for \Cyg\
  and (20:33:03.671208,+40:00:24.40818) for J2033+4000.   
  Column 4 gives  the peak brightness obtained for \Cyg\ by fitting a Gaussian distribution.
  Typical beam sizes were 0.5 mas FWHM.
  All errors are formal $1\sigma$ fitting uncertainties and do not include
  systematic errors.
               }
\label{table:cygx3_data}
\end{deluxetable}

The relative positions in Table \ref{table:cygx3_data} were modeled
by a trigonometric parallax signature and a linear proper motion and
fitted by variance-weighted least squares.  In order to account for
delay errors from uncompensated tropospheric water vapor, we added
``error floors'' in quadrature to the formal uncertainties in the East
and North offsets.  The error floors were adjusted to 
($\sigma_E,\sigma_N$) = ($\pm0.014,\pm0.070$) mas to give a reduced 
chi-squared per degree of freedom near unity in each coordinate. 
The reason for the error floor in the northerly direction, $\sigma_N$,
being five-times larger than in the easterly direction, $\sigma_E$,
is likely due to unresolved jitter in the core position owing to weak 
jetted emission in the North-South direction.

\begin{figure}[h]
\epsscale{0.85} 
\plotone{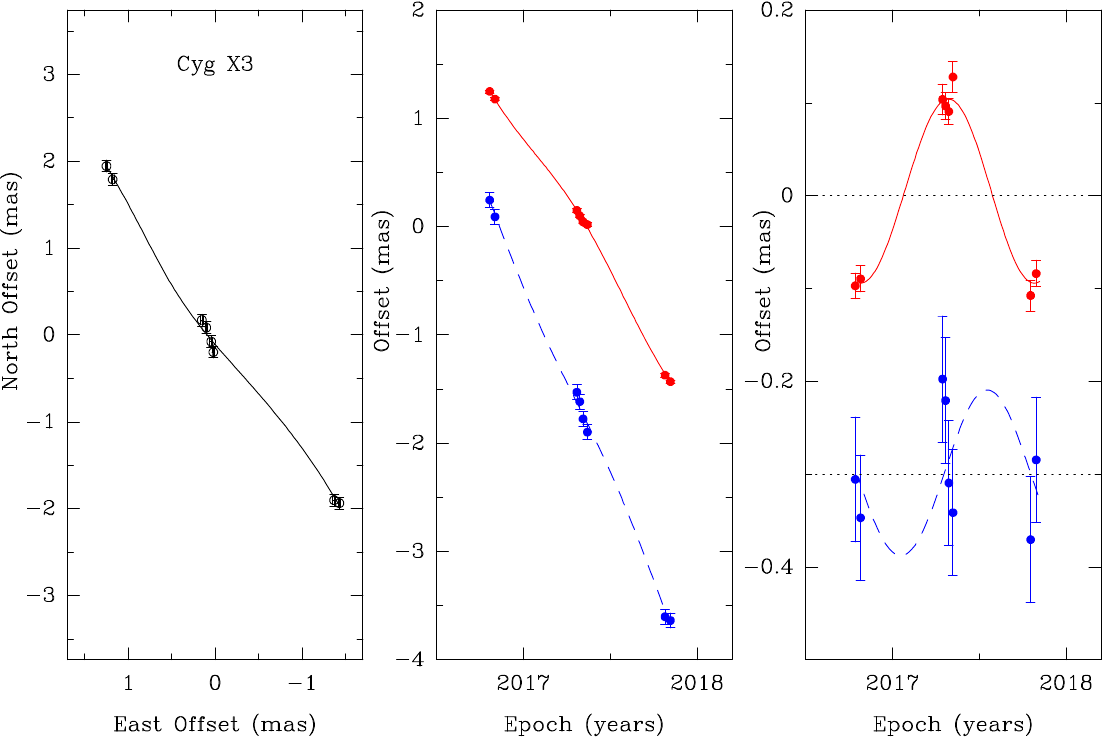}
\caption{\small Parallax data and fits for Cyg~X-3.  
{\it Left Panel:} Sky view with East and North offsets.
{\it Middle Panel:} East (data and fitted solid line in red) 
and North (data and fitted dashed line in blue) positions vs. time.
{\it Right Panel:} Same as middle panel, but with fitted proper motion
removed to highlight the parallax effect.  One-sigma error bars include 
systematic uncertainty added in quadrature with the formal fitting uncertainties,
yielding chi-squared per degree of freedom near unity in each coordinate.
}
\label{fig:parallax}
\end{figure}

Fig.~\ref{fig:parallax} displays the parallax data
and fits, as sky positions for all epochs, as well as East and North offsets
as a function of time.
The best-fit parallax is $0.1034\pm0.0054$ mas. With just a 5\% uncertainty, we can simply invert the parallax to determine the distance, without the need for a prior. This gives a distance of $9.67^{+0.53}_{-0.48}$ kpc.
The eastward and northward components of proper motion are $-2.589\pm0.014$
and $-3.747\pm0.069$ \masy.

\subsection{3D Kinematic Distance}
\label{sec:cyg_kinematics}

The proper motion of Cyg~X-3 had been measured relative to compact extragalactic
sources by \citet{Miller-Jones:09} to be $\ura=-2.73\pm0.06$ \masy\
y and $\udec=-3.70\pm0.06$ \masy\ using mostly Very Large Array A-configuration observations
spanning 1983 to 2006 at 8.4 GHz (or higher frequencies).  This motion is in reasonable
agreement with our more accurate measurement given above.

The line-of-sight velocity of this binary is very poorly constrained, owing to a
combination of high visual extinction and the infrared emission lines arising from differing locations in the turbulent wind of the Wolf-Rayet primary, which cannot therefore be used to determine the radial velocity of the system itself \citep[e.g.][]{Koljonen:17}.
Here we adopt a very loose prior on \vlsr\ of $-50 \pm 50$ \kms, which
is essentially consistent with it being a Galactic source toward a longitude of
$\approx80\deg$.

\begin{figure}
\epsscale{0.77} 
\plotone{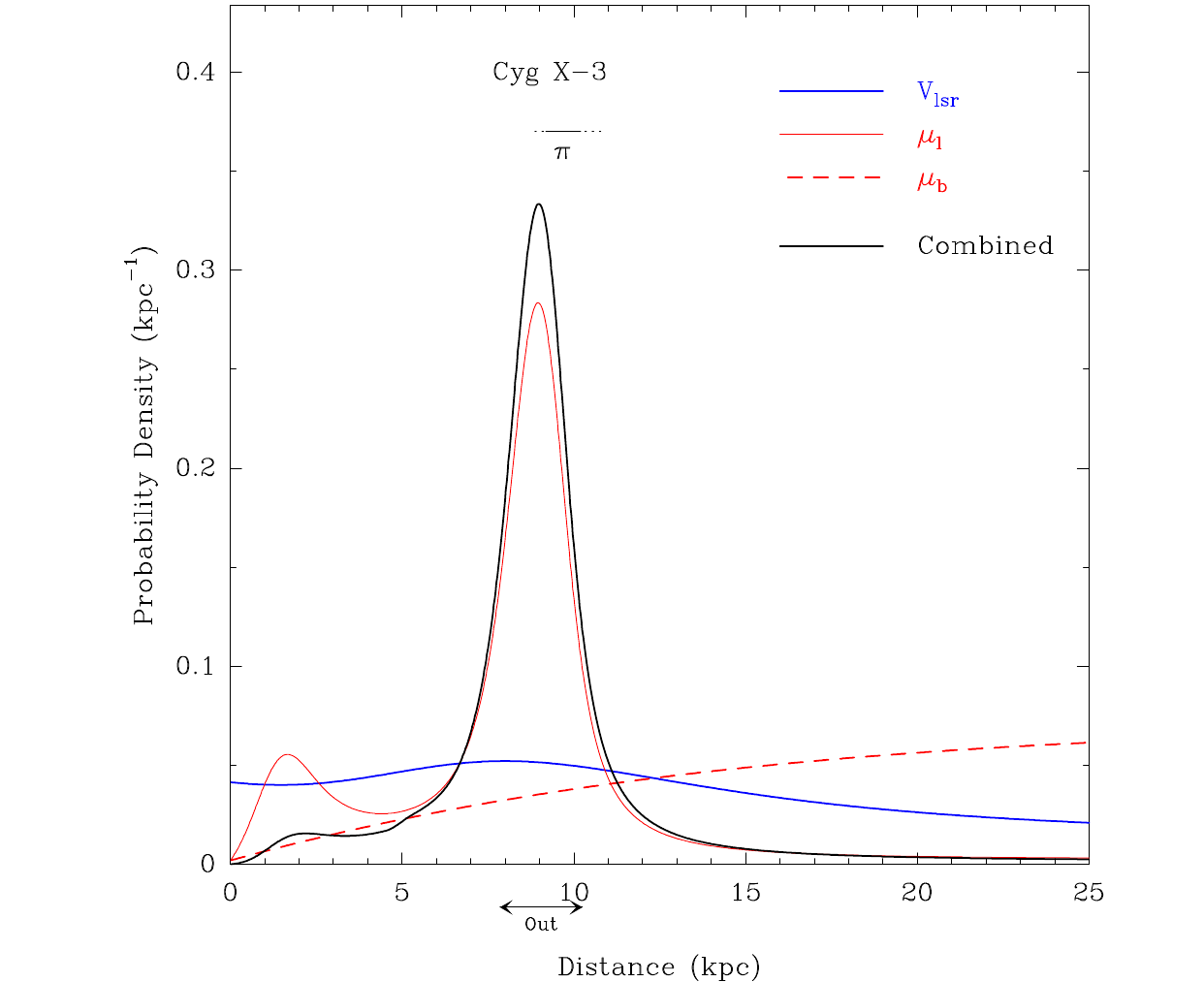}
\caption{\small Likelihoods for three components of motion, \vlsr\ in ({\it blue}),
  Galactic longitude ({\it red solid line}) and latitude ({\it red dashed line}),
  as a function of distance for Cyg~X-3.
  The product of the three likelihoods is shown in {\it black}, indicating
  a distance of $8.95 \pm 0.96$ kpc.
  The parallax distance reported in this paper is indicated with the $\pi$
  symbol, and its 68\% and 95\% confidence ranges are indicated with
  {\it solid} and {\it dotted lines} above it.  The distance range for the Outer
  spiral arm along the line-of-sight is indicated below the distance axis.
    }
\label{fig:CygX3}
\end{figure}

\begin{figure}
\epsscale{0.75} 
\plotone{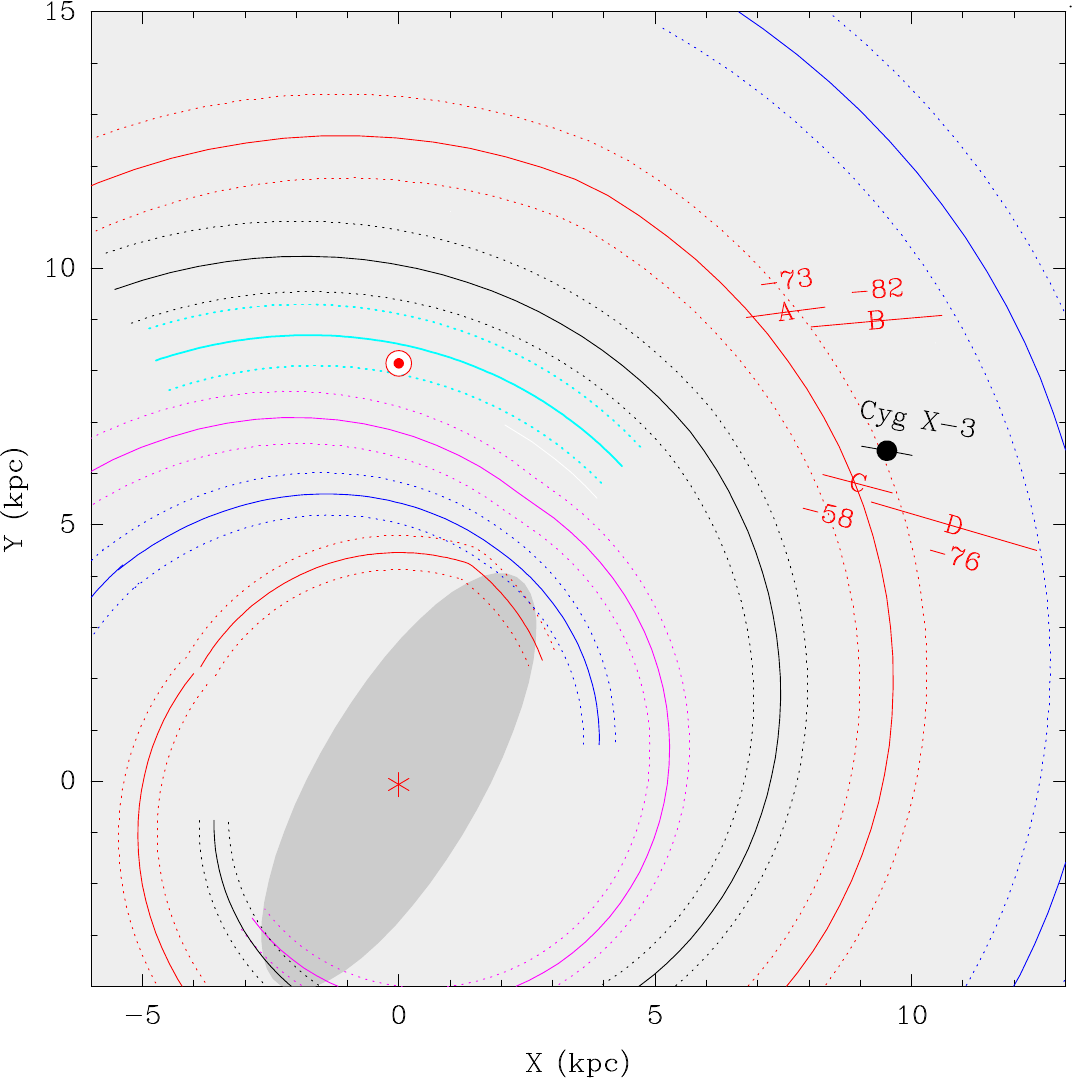}
\caption{\small Plan-view model of the Milky Way after \citet{Reid:19}, indicating
  the locations of \Cyg\ ({\it black dot}) and four massive, young stars with water
  masers ({\it red letters}) associated with the Outer spiral arm ({\it red lines})
  between Galactic longitudes 70\deg\ and 100\deg\ and with parallax distances
  from VLBI astrometry. A is G097.53$+$3.18 \citep{Hachisuka:15,Reid:19}, B is G095.05$+$3.97 \citep{Sakai:20}, C is G075.30$+$1.32 \citep{Sanna:12}, and D is G073.65$+$0.19 \citep{Reid:19}. \vlsr\ values in \kms\ for the masers
  are indicated next to their positions; each is uncertain by about $\pm10$ \kms.
    }
\label{fig:planview}
\end{figure}

Fig.~\ref{fig:CygX3} displays the likelihood functions for the three
components of motion of Cyg~X-3.  While the likelihood for its \vlsr\ provides
no useful constraint on distance, the proper motion component in Galactic longitude
strongly (and the latitude component weakly) constrains distance to be $8.95 \pm 0.96$ kpc.
Since the Cyg~X-3 binary contains a Wolf-Rayet star, which lives $\lax10$ Myr,
it should be very near its birth location inside a spiral arm of the Milky Way.
Fig.~\ref{fig:planview} shows the latest model of the spiral arms of the Milky Way
by \cite{Reid:19}, which is based on parallax measurements of
$\approx150$  massive, young stars.  The parallax (and kinematic) distance of 
Cyg~X-3 places it in the Outer spiral arm, which has a Galactic latitude of
$1\d7$ near the $79\d85$ longitude of Cyg~X-3.  The latitude of Cyg~X-3 differs
from that of other Outer arm by about 1\deg, corresponding to about 170 pc
at 9.67 kpc distance, which is well within the (Gaussian $1\sigma$) vertical
width of that arm of about 200 pc \citep[extrapolated from figure~4 of][]{Reid:19}.   

\begin{figure}
\epsscale{0.65} 
\plotone{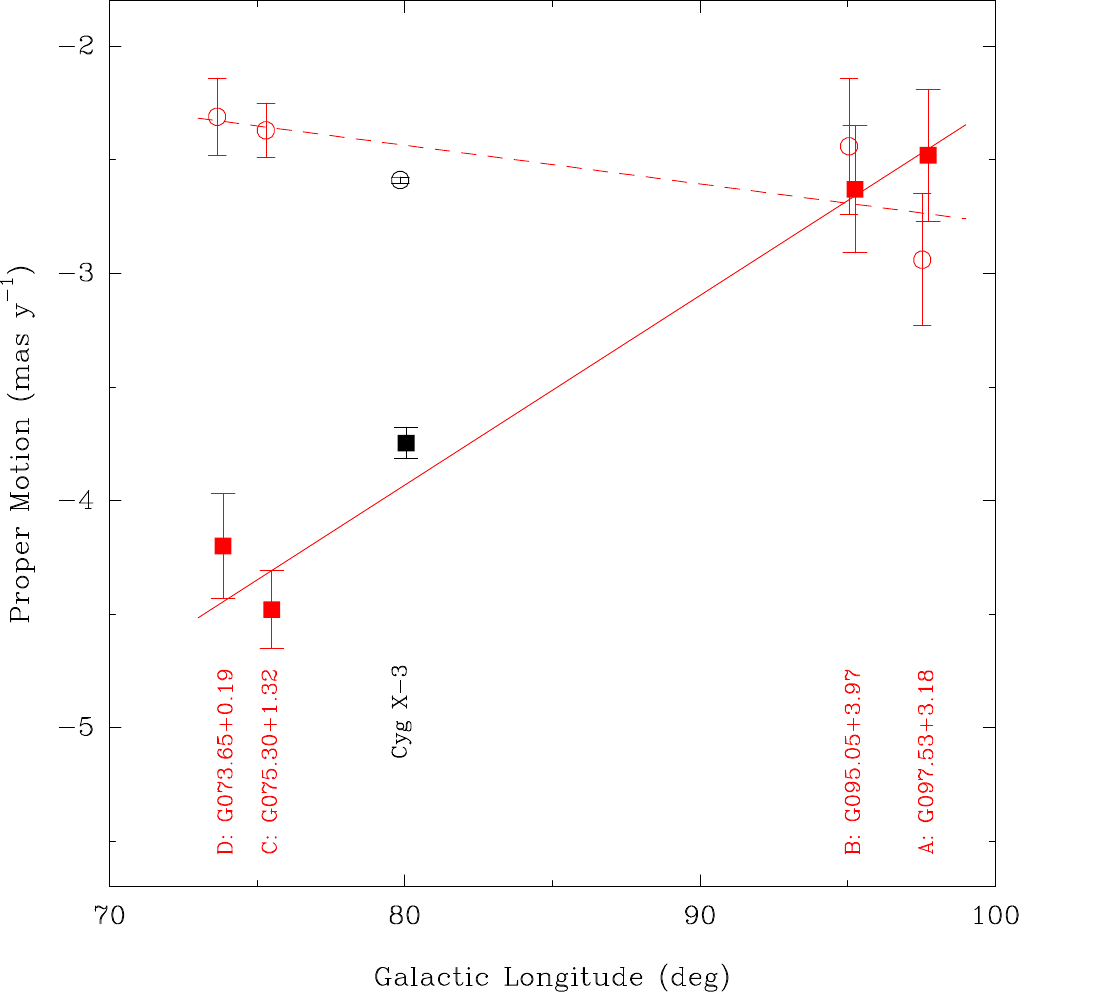}
\caption{\small Proper motions of four massive, young stars with water masers
  ({\it red symbols}) and Cyg~X-3 ({\it black symbols}) as a function of Galactic longitude.
  Source names are given below their measurements along with letter codes used in
  Fig. \ref{fig:planview} \citep{Sanna:12,Hachisuka:15,Reid:19,Sakai:20}.
  {\it Open circles} indicate motions in the easterly direction and {\it filled squares}
  indicate motions in the northerly direction.
  The young stars, which are associated with the Outer spiral arm of the Milky Way and
  straddle Cyg~X-3 in Galactic longitude, have motions consistent with that of Cyg~X-3.}
\label{fig:proper_motions}
\end{figure}

While \Cyg\ does not have a reliable line-of-sight velocity, its proper
motion has been accurately measured and can be compared to those of four massive,
young stars with maser astrometry that straddle \Cyg\ in Galactic longitude and are known to be in the Outer spiral arm, whose locations are shown in Fig.~\ref{fig:planview}.
Fig.~\ref{fig:proper_motions} shows the easterly
and northerly proper motions of these stars.  Note that both components of motion for \Cyg\
are consistent with interpolations between the sources, which 
bracket \Cyg\ in Galactic longitude.  This further supports the
association of Cyg~X-3 with the Outer arm.

\begin{figure}
\epsscale{0.65} 
\plotone{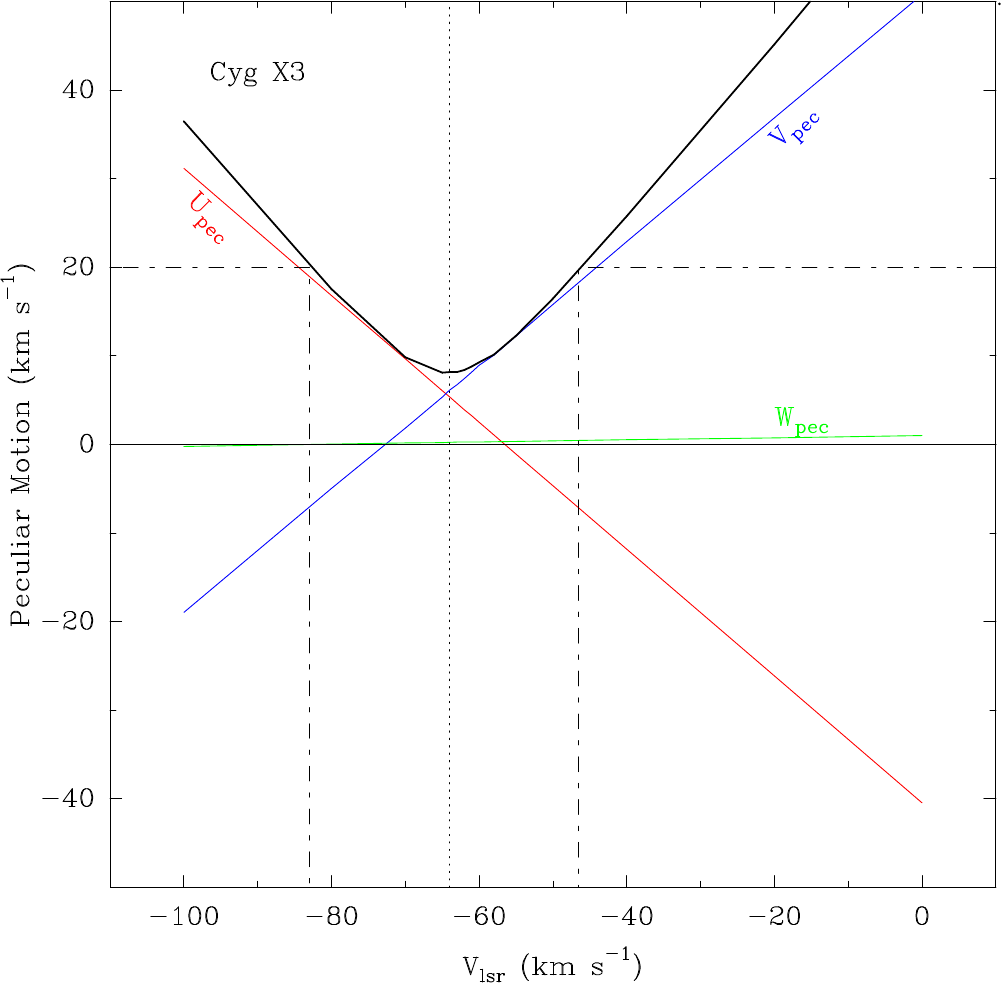}
\caption{\small Peculiar (non-circular) components of motion of Cyg~X-3 as a function
  of its (unknown) line-of-sight velocity component (\vlsr): $U_{pec}$ toward the
  Galactic center ({\it red}), $V_{pec}$ in the direction of Galactic rotation
  ({\it blue}), and $W_{pec}$ toward the North Galactic Pole ({\it green}).
  The 3D magnitude of the peculiar motion is plotted in {\it black} and has a minimum
  of 8 \kms\ at $\vlsr=-64$ \kms.   The \vlsr\ range for the motion magnitude being
  $>20$ \kms\ is shown by {\it dot-dashed lines}.
  Our parallax distance of 9.67 kpc is assumed.    
    }
\label{fig:peculiar_motion}
\end{figure}

Given the strong evidence that \Cyg\ formed recently in the
Outer spiral arm of the Milky Way, we now examine evidence that can constrain
its \vlsr.   The four massive, young stars which straddle Cyg~X-3 and have consistent
proper motions, have line-of-sight velocities that range from $-82 < \vlsr < -58$ \kms.
We now calculate the pecular motion of \Cyg\ as a function of
its (unknown) line-of-sight velocity and display this in Fig.~\ref{fig:peculiar_motion}.
The magnitude of the peculiar motion is less than 20 \kms\ for $-82 < \vlsr < -47$ \kms,
similar to the range of \vlsr\ for the four young stars, and there
is a clear minimum for the 3D peculiar motion of \Cyg\ near $\vlsr=-64$ \kms.
Note that our parallax distance and uncertainty, would yield standard (1D) kinematic
distance for \vlsr\ $=-64\pm5$ \kms. 

All together there is strong circumstantial evidence that Cyg~X-3 has
a small peculiar motion ($\lax20$ \kms), suggesting a small natal kick when
its compact star formed.   Indeed, the small value ($<1$ \kms) of the peculiar motion
component toward the North Galactic Pole, which is nearly independent of the \vlsr\
of Cyg~X-3, supports this conclusion, since it is {\it a priori} unlikely for
a natal kick to be entirely in the Galactic plane.

\section{Discussion}\label{discussion}

\subsection{\GRS}

The 3D kinematic distance estimate for \GRS\ is consistent with, but more precise than,
the parallax distance determined by \citet{Reid:14grs}. As noted in that work, the
distance affects the inferred black hole mass through its effect on the inclination of
the orbit (assumed to be aligned with the jets).  The proper motions $\mu_{\rm app}$
and $\mu_{\rm rec}$ of intrinsically symmetric approaching and receding jet ejecta can
be used to constrain the product $\beta \cos i$, where $\beta$ is the jet speed
normalized to the speed of light, and $i$ is the inclination angle of the jet axis to
the line of sight. With an accurate distance, these two parameters can be disentangled via
\begin{equation}
\tan i = \frac{2d}{c}\frac{\mu_{\rm app}\mu_{\rm rec}}{\mu_{\rm app}-\mu_{\rm rec}}.
\end{equation}

The larger distance of \GRS\ inferred from its 3D kinematics implies a larger inclination
angle for the jet axis. Using a census of paired ejecta with accurate proper motion
measurements from \citet{Miller-Jones:07}, we find a weighted mean inclination angle
of $64^{\circ}\pm4^{\circ}$. The increase in the inferred inclination angle implies a
higher jet speed for a given set of proper motion measurements, with the proper motions
of \citet{Miller-Jones:07} giving jet speeds ranging from 0.68--0.91$c$.

Given the $\sin^3 i$ dependence of the mass function on inclination, our higher
inclination \citep[relative to the $60^{\circ}\pm5^{\circ}$ determined from the parallax 
distance estimate by][]{Reid:14grs} would translate to a slight reduction in the inferred 
black hole mass, from 12.4\,$M_{\odot}$ to 11.2\,$M_{\odot}$. This reduction in inferred
black hole mass makes \GRS\ less of an outlier relative to the black hole mass
distribution of the low-mass X-ray binary population, as estimated by \citet{Farr:11}
and \citet{Kreidberg:12}.

As shown by \citet{Dhawan:07}, the peculiar velocity of \GRS\ is minimized at distances
of between 8 and 10\,kpc, such that our 3D kinematic distance does not significantly
impact the calculated non-circular motion of the system.  At only 36\,pc from the
Galactic plane, and with a peculiar velocity of $\sim20$\,km\,s$^{-1}$, \GRS\ is likely
to have formed either via direct collapse, or in a supernova with a very low natal kick.
Indeed, determining the potential kick velocity via the method of \citet{Atri:19},
we find a median of 32\,km\,s$^{-1}$, with a 90\% confidence interval of
17--65\,km\,s$^{-1}$. This is comparable to the lowest inferred natal kicks of any
low-mass X-ray binary.

\subsection{\Cyg}

Under the reasonable assumption (supported by strong circumstantial evidence, as detailed
in Section~\ref{sec:cyg_kinematics}) that the peculiar velocity of \Cyg\ is small, we
determine a 3D kinematic distance of $8.95\pm0.96$\,kpc, in good agreement with the
independently-determined trigonometric parallax measurement of $9.67^{+0.53}_{-0.48}$\,kpc.
This provides confidence in the distance determination and validates the effectiveness of
the 3D kinematic distance method \citep{Reid:22} for sources with a low peculiar velocity.

While early distance estimates for \Cyg\ \citep{Dickey:83,Predehl:00} placed the source
at a distance of $\sim10$\,kpc, more recent dust scattering measurements by \citet{Ling:09}
favored a lower distance of $7.2^{+0.3}_{-0.5}$\,kpc.  However, as noted by the authors,
this measurement was highly sensitive to the distance of the Cyg OB2 association
(assumed as 1.7\,kpc), and more recent {\it Gaia} data \citep{Berlanas:19} have shown the
cluster to be slightly more distant, at $\sim 1.76$\,kpc.  Extended X-ray emission which
varied on the orbital period of \Cyg\ was found to arise from X-ray scattering by a Bok
globule along the line of sight \citep{McCollough:13}.  Standard 1D kinematic distances
to the globule with $\vlsr=-47.5$ \kms\ are either $6.1\pm0.6$ or $7.8\pm0.6$ kpc.
Modeling the time delay of the scattered light curve yielded possible distances to
Cyg~X-3 of either $7.4\pm1.1$ and $10.2\pm1.2$ kpc, at 62 and 38\% probability,
respectively \citep{McCollough:16}. The farther distance estimate would be fully
consistent with our measurement.

Since our distance measurement is consistent with that of earlier works
\citep{Dickey:83,Predehl:00}, the update does not significantly change jet velocities
calculated by \citet{Mioduszewski:01} or \citet{Miller-Jones:04}.  However,
\citet{Koljonen:17} noted that a distance of $\sim10$\,kpc would imply a slight
increase in the inferred mass of the Wolf-Rayet donor, to a range of 11--14\,$M_{\odot}$.

Recent X-ray polarization measurements from the Imaging X-ray Polarimetry Explorer
(IXPE) suggested that the central compact object in Cyg X-3 is highly obscured
\citep{Veledina:23}, likely due to an optically-thick envelope which surrounds a narrow
funnel, whose walls allow reflected and scattered light to escape.  Unless the opening
angle of the funnel was very small ($\lesssim16^{\circ}$), the inferred geometry would
suggest an intrinsic luminosity exceeding the Eddington limit, even for an accretor in
excess of $20M_{\odot}$.  Since these calculations were based on the lower distance of
\citet{Ling:09}, these inclination angle limits should be even more stringent,
strengthening the argument that Cyg X-3 would be an ultraluminous X-ray source if
observed face-on.

Furthermore, recent work by \citet{Koljonen:23} suggested a spatial and temporal
association between gamma-ray flaring in Cyg X-3 and IceCube neutrino detections,
raising the possibility that protons could be accelerated to highly-relativistic
energies within the jets of this system, and making Cyg X-3 a possible source of
cosmic rays.  Our new distance determination would allow for a more accurate assessment
of the proton luminosity, and hence the potential cosmic ray contribution of
microquasar systems.

\section{Conclusions and Outlook}\label{conclusions}

X-ray binary distance measurements are crucial to understanding their nature,
allowing us to determine their underlying physical parameters. For \Cyg, we
have measured a highly accurate trigonometric parallax distance of
$9.67^{+0.53}_{-0.48}$\,kpc.  To date, this is the most distant X-ray binary
with a radio parallax measurement, demonstrating the potential of VLBI measured
parallaxes for bright sources, even at large distances along highly scatter-broadened
lines-of-sight.  Our refined distance measurement strengthens the argument that
\Cyg\ would appear as an ultraluminous X-ray source if viewed face-on.

Using \Cyg's measured proper motion, we determine a 3D kinematic distance of
$8.95\pm0.96$\,kpc, which is consistent with the more accurate parallax distance,
demonstrating that 3D kinematics can provide reliable distance estimates for
X-ray binaries with low peculiar velocities.  Both the parallax and kinematic
distance locate the system within the Outer spiral arm of the Galaxy.
Its proper motion is consistent with those of young, massive stars in the same
region, which have LSR velocities near $-70$ \kms, suggesting that \Cyg\ has a
similar LSR velocity and supporting a small peculiar velocity of $<20$ \kms. 

We also estimated a 3D kinematic distance for \GRS\ of
$9.4\pm0.6~ (\rm{stat.}) \pm0.8~ (\rm{sys.})$\,kpc.  This distance is consistent
with, but more precise than the previous parallax result of \citet{Reid:14grs}.
At 9.4 kpc, \GRS\ would have a slightly higher inclination angle, and hence
lower black hole mass, than previously suggested.

This work underscores the importance of high-precision astrometric measurements
of X-ray binary systems, even in cases where a parallax distance measurement is
not possible (e.g.,\ due to the lack of a close calibrator, a large distance,
or significant scatter broadening).  When coupled with a line-of-sight radial
velocity measurement, they can provide reliable 3D kinematic distances for
sources with low peculiar velocities.

\begin{acknowledgments}
We thank Arash Bahramian for useful discussions and assistance with fitting jet
parameters. This work made use of the Swinburne University of Technology
software correlator, developed as part of the Australian Major National Research
Facilities Programme and operated under licence.
This work has made use of NASA's Astrophysics Data System.
\end{acknowledgments}

\vspace{5mm}
\facilities{VLBA}


\software{AIPS \citep{Greisen:2003}
          }

\bibliography{cygx3_grs1915}{}

\begin{thebibliography}{}
\expandafter\ifx\csname natexlab\endcsname\relax\def\natexlab#1{#1}\fi
\providecommand{\url}[1]{\href{#1}{#1}}
\providecommand{\dodoi}[1]{doi:~\href{http://doi.org/#1}{\nolinkurl{#1}}}
\providecommand{\doeprint}[1]{\href{http://ascl.net/#1}{\nolinkurl{http://ascl.net/#1}}}
\providecommand{\doarXiv}[1]{\href{https://arxiv.org/abs/#1}{\nolinkurl{https://arxiv.org/abs/#1}}}

\bibitem[{{Atri} {et~al.}(2019){Atri}, {Miller-Jones}, {Bahramian}, {Plotkin},
  {Jonker}, {Nelemans}, {Maccarone}, {Sivakoff}, {Deller}, {Chaty}, {Torres},
  {Horiuchi}, {McCallum}, {Natusch}, {Phillips}, {Stevens}, \&
  {Weston}}]{Atri:19}
{Atri}, P., {Miller-Jones}, J.~C.~A., {Bahramian}, A., {et~al.} 2019, \mnras,
  489, 3116, \dodoi{10.1093/mnras/stz2335}

\bibitem[{{Berlanas} {et~al.}(2019){Berlanas}, {Wright}, {Herrero}, {Drew}, \&
  {Lennon}}]{Berlanas:19}
{Berlanas}, S.~R., {Wright}, N.~J., {Herrero}, A., {Drew}, J.~E., \& {Lennon},
  D.~J. 2019, \mnras, 484, 1838, \dodoi{10.1093/mnras/stz117}

\bibitem[{{Bolton}(1972)}]{Bolton:72}
{Bolton}, C.~T. 1972, \nat, 235, 271, \dodoi{10.1038/235271b0}

\bibitem[{{Caballero-Nieves} {et~al.}(2009){Caballero-Nieves}, {Gies},
  {Bolton}, {Hadrava}, {Herrero}, {Hillwig}, {Howell}, {Huang}, {Kaper},
  {Koubsk{\'y}}, \& {McSwain}}]{Caballero-Nieves:09}
{Caballero-Nieves}, S.~M., {Gies}, D.~R., {Bolton}, C.~T., {et~al.} 2009, \apj,
  701, 1895, \dodoi{10.1088/0004-637X/701/2/1895}

\bibitem[{{Deller} {et~al.}(2011){Deller}, {Brisken}, {Phillips}, {Morgan},
  {Alef}, {Cappallo}, {Middelberg}, {Romney}, {Rottmann}, {Tingay}, \&
  {Wayth}}]{Deller:11}
{Deller}, A.~T., {Brisken}, W.~F., {Phillips}, C.~J., {et~al.} 2011, \pasp,
  123, 275, \dodoi{10.1086/658907}

\bibitem[{{Dhawan} {et~al.}(2007){Dhawan}, {Mirabel}, {Rib{\'o}}, \&
  {Rodrigues}}]{Dhawan:07}
{Dhawan}, V., {Mirabel}, I.~F., {Rib{\'o}}, M., \& {Rodrigues}, I. 2007, \apj,
  668, 430, \dodoi{10.1086/520111}

\bibitem[{{Dickey}(1983)}]{Dickey:83}
{Dickey}, J.~M. 1983, \apjl, 273, L71, \dodoi{10.1086/184132}

\bibitem[{{Do} {et~al.}(2019){Do}, {Hees}, {Ghez}, {Martinez}, {Chu}, {Jia},
  {Sakai}, {Lu}, {Gautam}, {O'Neil}, {Becklin}, {Morris}, {Matthews},
  {Nishiyama}, {Campbell}, {Chappell}, {Chen}, {Ciurlo}, {Dehghanfar},
  {Gallego-Cano}, {Kerzendorf}, {Lyke}, {Naoz}, {Saida}, {Sch{\"o}del},
  {Takahashi}, {Takamori}, {Witzel}, \& {Wizinowich}}]{2019Sci...365..664D}
{Do}, T., {Hees}, A., {Ghez}, A., {et~al.} 2019, Science, 365, 664,
  \dodoi{10.1126/science.aav8137}

\bibitem[{{Farr} {et~al.}(2011){Farr}, {Sravan}, {Cantrell}, {Kreidberg},
  {Bailyn}, {Mandel}, \& {Kalogera}}]{Farr:11}
{Farr}, W.~M., {Sravan}, N., {Cantrell}, A., {et~al.} 2011, \apj, 741, 103,
  \dodoi{10.1088/0004-637X/741/2/103}

\bibitem[{{Fender} {et~al.}(1999){Fender}, {Garrington}, {McKay}, {Muxlow},
  {Pooley}, {Spencer}, {Stirling}, \& {Waltman}}]{Fender:99}
{Fender}, R.~P., {Garrington}, S.~T., {McKay}, D.~J., {et~al.} 1999, \mnras,
  304, 865, \dodoi{10.1046/j.1365-8711.1999.02364.x}

\bibitem[{{GRAVITY Collaboration} {et~al.}(2021){GRAVITY Collaboration},
  {Abuter}, {Amorim}, {Baub{\"o}ck}, {Berger}, {Bonnet}, {Brandner},
  {Cl{\'e}net}, {Davies}, {de Zeeuw}, {Dexter}, {Dallilar}, {Drescher},
  {Eckart}, {Eisenhauer}, {F{\"o}rster Schreiber}, {Garcia}, {Gao}, {Gendron},
  {Genzel}, {Gillessen}, {Habibi}, {Haubois}, {Hei{\ss}el}, {Henning},
  {Hippler}, {Horrobin}, {Jim{\'e}nez-Rosales}, {Jochum}, {Jocou}, {Kaufer},
  {Kervella}, {Lacour}, {Lapeyr{\`e}re}, {Le Bouquin}, {L{\'e}na}, {Lutz},
  {Nowak}, {Ott}, {Paumard}, {Perraut}, {Perrin}, {Pfuhl}, {Rabien},
  {Rodr{\'\i}guez-Coira}, {Shangguan}, {Shimizu}, {Scheithauer}, {Stadler},
  {Straub}, {Straubmeier}, {Sturm}, {Tacconi}, {Vincent}, {von Fellenberg},
  {Waisberg}, {Widmann}, {Wieprecht}, {Wiezorrek}, {Woillez}, {Yazici},
  {Young}, \& {Zins}}]{2021A&A...647A..59G}
{GRAVITY Collaboration}, {Abuter}, R., {Amorim}, A., {et~al.} 2021, \aap, 647,
  A59, \dodoi{10.1051/0004-6361/202040208}

\bibitem[{{Greisen}(2003)}]{Greisen:2003}
{Greisen}, E.~W. 2003, in Astrophysics and Space Science Library, Vol. 285,
  Information Handling in Astronomy - Historical Vistas, ed. A.~{Heck}, 109,
  \dodoi{10.1007/0-306-48080-8_7}

\bibitem[{{Hachisuka} {et~al.}(2015){Hachisuka}, {Choi}, {Reid}, {Brunthaler},
  {Menten}, {Sanna}, \& {Dame}}]{Hachisuka:15}
{Hachisuka}, K., {Choi}, Y.~K., {Reid}, M.~J., {et~al.} 2015, \apj, 800, 2,
  \dodoi{10.1088/0004-637X/800/1/2}

\bibitem[{{Koljonen} \& {Maccarone}(2017)}]{Koljonen:17}
{Koljonen}, K.~I.~I., \& {Maccarone}, T.~J. 2017, \mnras, 472, 2181,
  \dodoi{10.1093/mnras/stx2106}

\bibitem[{{Koljonen} {et~al.}(2023){Koljonen}, {Satalecka}, {Lindfors}, \&
  {Liodakis}}]{Koljonen:23}
{Koljonen}, K.~I.~I., {Satalecka}, K., {Lindfors}, E.~J., \& {Liodakis}, I.
  2023, \mnras, \dodoi{10.1093/mnrasl/slad081}

\bibitem[{{Kreidberg} {et~al.}(2012){Kreidberg}, {Bailyn}, {Farr}, \&
  {Kalogera}}]{Kreidberg:12}
{Kreidberg}, L., {Bailyn}, C.~D., {Farr}, W.~M., \& {Kalogera}, V. 2012, \apj,
  757, 36, \dodoi{10.1088/0004-637X/757/1/36}

\bibitem[{{Ling} {et~al.}(2009){Ling}, {Zhang}, \& {Tang}}]{Ling:09}
{Ling}, Z., {Zhang}, S.~N., \& {Tang}, S. 2009, \apj, 695, 1111,
  \dodoi{10.1088/0004-637X/695/2/1111}

\bibitem[{{McCollough} {et~al.}(2016){McCollough}, {Corrales}, \&
  {Dunham}}]{McCollough:16}
{McCollough}, M.~L., {Corrales}, L., \& {Dunham}, M.~M. 2016, \apjl, 830, L36,
  \dodoi{10.3847/2041-8205/830/2/L36}

\bibitem[{{McCollough} {et~al.}(2013){McCollough}, {Smith}, \&
  {Valencic}}]{McCollough:13}
{McCollough}, M.~L., {Smith}, R.~K., \& {Valencic}, L.~A. 2013, \apj, 762, 2,
  \dodoi{10.1088/0004-637X/762/1/2}

\bibitem[{{Miller-Jones} {et~al.}(2004){Miller-Jones}, {Blundell}, {Rupen},
  {Mioduszewski}, {Duffy}, \& {Beasley}}]{Miller-Jones:04}
{Miller-Jones}, J. C.~A., {Blundell}, K.~M., {Rupen}, M.~P., {et~al.} 2004,
  \apj, 600, 368, \dodoi{10.1086/379706}

\bibitem[{{Miller-Jones} {et~al.}(2007){Miller-Jones}, {Rupen}, {Fender},
  {Rushton}, {Pooley}, \& {Spencer}}]{Miller-Jones:07}
{Miller-Jones}, J.~C.~A., {Rupen}, M.~P., {Fender}, R.~P., {et~al.} 2007,
  \mnras, 375, 1087, \dodoi{10.1111/j.1365-2966.2007.11381.x}

\bibitem[{{Miller-Jones} {et~al.}(2009){Miller-Jones}, {Sakari}, {Dhawan},
  {Tudose}, {Fender}, {Paragi}, \& {Garrett}}]{Miller-Jones:09}
{Miller-Jones}, J.~C.~A., {Sakari}, C.~M., {Dhawan}, V., {et~al.} 2009, in 8th
  International e-VLBI Workshop, 17.
\newblock \doarXiv{0909.2589}

\bibitem[{{Miller-Jones} {et~al.}(2021){Miller-Jones}, {Bahramian}, {Orosz},
  {Mandel}, {Gou}, {Maccarone}, {Neijssel}, {Zhao}, {Zi{\'o}{\l}kowski},
  {Reid}, {Uttley}, {Zheng}, {Byun}, {Dodson}, {Grinberg}, {Jung}, {Kim},
  {Marcote}, {Markoff}, {Rioja}, {Rushton}, {Russell}, {Sivakoff}, {Tetarenko},
  {Tudose}, \& {Wilms}}]{Miller-Jones:21}
{Miller-Jones}, J. C.~A., {Bahramian}, A., {Orosz}, J.~A., {et~al.} 2021,
  Science, 371, 1046, \dodoi{10.1126/science.abb3363}

\bibitem[{{Mioduszewski} {et~al.}(2001){Mioduszewski}, {Rupen}, {Hjellming},
  {Pooley}, \& {Waltman}}]{Mioduszewski:01}
{Mioduszewski}, A.~J., {Rupen}, M.~P., {Hjellming}, R.~M., {Pooley}, G.~G., \&
  {Waltman}, E.~B. 2001, \apj, 553, 766, \dodoi{10.1086/320965}

\bibitem[{{Mirabel} \& {Rodr{\'\i}guez}(1994)}]{Mirabel:94}
{Mirabel}, I.~F., \& {Rodr{\'\i}guez}, L.~F. 1994, \nat, 371, 46,
  \dodoi{10.1038/371046a0}

\bibitem[{{Persic} {et~al.}(1996){Persic}, {Salucci}, \& {Stel}}]{Persic:96}
{Persic}, M., {Salucci}, P., \& {Stel}, F. 1996, \mnras, 281, 27,
  \dodoi{10.1093/mnras/278.1.27}

\bibitem[{{Predehl} {et~al.}(2000){Predehl}, {Burwitz}, {Paerels}, \&
  {Tr{\"u}mper}}]{Predehl:00}
{Predehl}, P., {Burwitz}, V., {Paerels}, F., \& {Tr{\"u}mper}, J. 2000, \aap,
  357, L25

\bibitem[{{Reid}(2022)}]{Reid:22}
{Reid}, M.~J. 2022, \aj, 164, 133, \dodoi{10.3847/1538-3881/ac80bb}

\bibitem[{{Reid} \& {Brunthaler}(2020)}]{2020ApJ...892...39R}
{Reid}, M.~J., \& {Brunthaler}, A. 2020, \apj, 892, 39,
  \dodoi{10.3847/1538-4357/ab76cd}

\bibitem[{{Reid} {et~al.}(2011){Reid}, {McClintock}, {Narayan}, {Gou},
  {Remillard}, \& {Orosz}}]{Reid:11}
{Reid}, M.~J., {McClintock}, J.~E., {Narayan}, R., {et~al.} 2011, \apj, 742,
  83, \dodoi{10.1088/0004-637X/742/2/83}

\bibitem[{{Reid} {et~al.}(2014){Reid}, {McClintock}, {Steiner}, {Steeghs},
  {Remillard}, {Dhawan}, \& {Narayan}}]{Reid:14grs}
{Reid}, M.~J., {McClintock}, J.~E., {Steiner}, J.~F., {et~al.} 2014, \apj, 796,
  2, \dodoi{10.1088/0004-637X/796/1/2}

\bibitem[{{Reid} {et~al.}(2009){Reid}, {Menten}, {Brunthaler}, {Zheng},
  {Moscadelli}, \& {Xu}}]{Reid:09}
{Reid}, M.~J., {Menten}, K.~M., {Brunthaler}, A., {et~al.} 2009, \apj, 693,
  397, \dodoi{10.1088/0004-637X/693/1/397}

\bibitem[{{Reid} {et~al.}(2019){Reid}, {Menten}, {Brunthaler}, {Zheng}, {Dame},
  {Xu}, {Li}, {Sakai}, {Wu}, {Immer}, {Zhang}, {Sanna}, {Moscadelli}, {Rygl},
  {Bartkiewicz}, {Hu}, {Quiroga-Nu{\~n}ez}, \& {van Langevelde}}]{Reid:19}
---. 2019, \apj, 885, 131, \dodoi{10.3847/1538-4357/ab4a11}

\bibitem[{{Sakai} {et~al.}(2020){Sakai}, {Nagayama}, {Nakanishi}, {Koide},
  {Kurayama}, {Izumi}, {Hirota}, {Yoshida}, {Shibata}, \& {Honma}}]{Sakai:20}
{Sakai}, N., {Nagayama}, T., {Nakanishi}, H., {et~al.} 2020, \pasj, 72, 53,
  \dodoi{10.1093/pasj/psz125}

\bibitem[{{Sanna} {et~al.}(2012){Sanna}, {Reid}, {Dame}, {Menten},
  {Brunthaler}, {Moscadelli}, {Zheng}, \& {Xu}}]{Sanna:12}
{Sanna}, A., {Reid}, M.~J., {Dame}, T.~M., {et~al.} 2012, \apj, 745, 82,
  \dodoi{10.1088/0004-637X/745/1/82}

\bibitem[{{Steeghs} {et~al.}(2013){Steeghs}, {McClintock}, {Parsons}, {Reid},
  {Littlefair}, \& {Dhillon}}]{Steeghs:13}
{Steeghs}, D., {McClintock}, J.~E., {Parsons}, S.~G., {et~al.} 2013, \apj, 768,
  185, \dodoi{10.1088/0004-637X/768/2/185}

\bibitem[{{Veledina} {et~al.}(2023){Veledina}, {Muleri}, {Poutanen},
  {Podgorn{\'y}}, {Dov{\v{c}}iak}, {Capitanio}, {Churazov}, {De Rosa}, {Di
  Marco}, {Forsblom}, {Kaaret}, {Krawczynski}, {La Monaca}, {Loktev},
  {Lutovinov}, {Molkov}, {Mushtukov}, {Ratheesh}, {Rodriguez Cavero},
  {Steiner}, {Sunyaev}, {Tsygankov}, {Zdziarski}, {Bianchi}, {Bright},
  {Bursov}, {Costa}, {Egron}, {Garcia}, {Green}, {Gurwell}, {Ingram}, {Kajava},
  {Kale}, {Kraus}, {Malyshev}, {Marin}, {Matt}, {McCollough}, {Mereminskiy},
  {Nizhelsky}, {Piano}, {Pilia}, {Pittori}, {Rao}, {Righini}, {Soffitta},
  {Shevchenko}, {Svoboda}, {Tombesi}, {Trushkin}, {Tsybulev}, {Ursini},
  {Weisskopf}, {Wu}, {Agudo}, {Antonelli}, {Bachetti}, {Baldini},
  {Baumgartner}, {Bellazzini}, {Bongiorno}, {Bonino}, {Brez}, {Bucciantini},
  {Castellano}, {Cavazzuti}, {Chen}, {Ciprini}, {Del Monte}, {Di Gesu}, {Di
  Lalla}, {Donnarumma}, {Doroshenko}, {Ehlert}, {Enoto}, {Evangelista},
  {Fabiani}, {Ferrazzoli}, {Gunji}, {Hayashida}, {Heyl}, {Iwakiri}, {Jorstad},
  {Karas}, {Kislat}, {Kitaguchi}, {Kolodziejczak}, {Latronico}, {Liodakis},
  {Maldera}, {Manfreda}, {Marinucci}, {Marscher}, {Marshall}, {Massaro},
  {Mitsuishi}, {Mizuno}, {Negro}, {Ng}, {O'Dell}, {Omodei}, {Oppedisano},
  {Papitto}, {Pavlov}, {Peirson}, {Perri}, {Pesce-Rollins}, {Petrucci},
  {Possenti}, {Puccetti}, {Ramsey}, {Rankin}, {Roberts}, {Romani}, {Sgr{\`o}},
  {Slane}, {Spandre}, {Swartz}, {Tamagawa}, {Tavecchio}, {Taverna}, {Tawara},
  {Tennant}, {Thomas}, {Trois}, {Turolla}, {Vink}, {Xie}, \&
  {Zane}}]{Veledina:23}
{Veledina}, A., {Muleri}, F., {Poutanen}, J., {et~al.} 2023, arXiv e-prints,
  arXiv:2303.01174, \dodoi{10.48550/arXiv.2303.01174}

\bibitem[{{Webster} \& {Murdin}(1972)}]{Webster:72}
{Webster}, B.~L., \& {Murdin}, P. 1972, \nat, 235, 37, \dodoi{10.1038/235037a0}

\bibitem[{{Wu} {et~al.}(2014){Wu}, {Sato}, {Reid}, {Moscadelli}, {Zhang}, {Xu},
  {Brunthaler}, {Menten}, {Dame}, \& {Zheng}}]{Wu:14}
{Wu}, Y.~W., {Sato}, M., {Reid}, M.~J., {et~al.} 2014, \aap, 566, A17,
  \dodoi{10.1051/0004-6361/201322765}

\end{thebibliography}
\bibliographystyle{aasjournal}

\end{document}